\documentclass[amsmath,amssymb,aps,preprint]{revtex4}

\usepackage{graphicx}
\usepackage{dcolumn}
\usepackage{bm}
\begin{document}
\title{Weak Quasi-elastic Production of Hyperons}
\author{S. K. Singh$^{a,b}$ and M. J. Vicente Vacas$^{a}$}
\affiliation{$^{a}$Departamento de F\'{\i}sica Te\'orica and IFIC, 
Centro Mixto Universidad de Valencia CSIC;
Institutos de Investigaci\'on de Paterna, Aptdo. 22085, 46071 Valencia, Spain.\\
$^b$Department of Physics, Aligarh Muslim University, Aligarh- 202002, India.}
       
\email{pht13sks@rediffmail.com}

\date{\today}

\begin{abstract}
The quasielastic weak production of $\Lambda$ and $\Sigma$ hyperons
from nucleons and nuclei induced by antineutrinos is studied in the energy region of
some ongoing neutrino oscillation experiments in the intermediate energy region. The hyperon
nucleon transition form factors determined from neutrino nucleon scattering and an analysis of high
precision data on semileptonic decays of neutron and hyperons using SU(3)
symmetry have been used. The nuclear effects due to  Fermi motion and final state
interaction  effects due to hyperon nucleon scattering have also been studied. The
numerical results for  differential and total cross sections have
been presented.
\end{abstract}

\maketitle 

\section{Introduction}
                 
The study of weak nuclear reactions induced by neutrinos and antineutrinos in the energy
region of few GeV has become quite important due to the role played by these processes in
the analysis of various neutrino oscillation experiments being done with atmospheric and
accelerator neutrinos in the intermediate energy 
region~\cite{Fukuda:1998mi,Ahn:2002up,Monroe:2004su,Adamson:2005qc}. In this energy region,
the theoretical cross sections for various weak processes induced by neutrinos and
antineutrinos on nucleons and nuclei are needed to model neutrino-nuclear interactions in
Monte Carlo neutrino generators like NUANCE~\cite{Casper:2002sd},
NEUGEN~\cite{Gallagher:2002sf}, NEUT~\cite{neut} or more general codes like
FLUKA~\cite{fluka} which are being used by groups  doing neutrino oscillation experiments.
The dominant weak process of current interest is the quasi-elastic
production of leptons induced by $\Delta S=0$  charged and neutral weak currents which has
been extensively studied in literature including  nuclear effects using various approaches
~\cite{Smith:1972xh,Gaisser:1986bv,Kuramoto:1989tk,Singh:1992dc,Mann:1993qp,
Umino:1996cz,Kim:1994ze,Co':2002kj,Nakamura:2002sg,Benhar:2005dj,Meucci:2004ip,
Nieves:2004wx,Nieves:2005rq,Martinez:2005xe}. However, in this energy region other
processes in which pions, kaons and hyperons are produced can also be important. In 
particular, the inelastic processes where  single pions are produced by weak charged and
neutral currents have recently  attracted much attention as they play a very important
role in performing the background studies in the analysis of neutrino oscillation
experiments. Many authors ~\cite{paschos,sato,Kim:1996bt,Singh:1998ha,Marteau:1999jp,
Paschos:2003qr,Leitner:2006ww,Cassing:2006sg} have recently studied the weak pion
production from nucleons and nuclei in the energy region relevant for the ongoing neutrino
oscillation experiments by K2K\cite{Ahn:2002up} and MiniBooNE collaborations\cite{Monroe:2004su}. 
In some
of these studies the nuclear effects in the weak pion production process as well as in the
final state interaction (FSI) of outgoing pions with the final nucleus have  also been taken
into account ~\cite{Kim:1996bt,Singh:1998ha,Marteau:1999jp,Paschos:2003qr,Leitner:2006ww,
Cassing:2006sg}.

 There exist very few calculations for the neutrino production of strange baryons and
mesons from free nucleons. In these calculations the hyperon nucleon transition form
factors are determined either from the Cabibbo theory with SU(3) symmetry ~\cite{cab1,cab2} or
from some quark models used for describing the baryon structure\cite{frinjord}. There are no 
calculations
to our knowledge where nuclear effects have been included in the weak 
production of strange particles from nuclei induced by neutrinos. The neutrino production
of strange particles is induced by weak charged as well as neutral currents. The weak
neutral currents induce only $\Delta S=0$ processes due to absence of Flavour Changing
Neutral Currents (in the standard model). On the other hand, the weak charged
currents induce both $\Delta S=0$ and $\Delta S=1$ processes. The production of strange
particles through $\Delta S=1$ processes is suppressed by a factor $tan^2\theta_c$ where
$\theta_c$ is the Cabibbo angle, as compared to the $\Delta S=0$ processes. However, in the
low energy region of $E_\nu\sim 1-3$GeV, the associated production of strange particles
through $\Delta S=0$ processes is suppressed by phase space. Therefore, it is likely that
in this low energy region, the cross sections for the production of strange particles
through $\Delta S=1$ and $\Delta S=0$ processes become comparable. In the case of the weak
production of strange particles through $\Delta S=1$ processes, the $\Delta S=\Delta Q$
selection rule restricts the quasi elastic hyperon production to antineutrinos rather than
neutrinos. As a consequence, in the $\Delta S= 1$ sector only antineutrino induced reactions
like ${\bar{\nu_l}} +N\rightarrow l^+ + Y(Y^\star)$ where $Y(Y^\star)$ is a $S=-1$ hyperon
(hyperon resonance) are allowed. Therefore, the  only possible quasi-elastic $\Delta S=1$
hyperon $(Y)$ production processes allowed in the neutrino(antineutrino) induced reactions are
\begin{eqnarray} 
{\bar{\nu_l}} +p&\rightarrow l^+ +\Lambda\\ 
{\bar{\nu_l}} +p&\rightarrow l^+ + \Sigma^0\\ 
{\bar{\nu_l}} +n&\rightarrow l^+ +\Sigma^- 
\end{eqnarray} 
These reactions have been experimentally studied in past but the
experimental information is very scanty and comes mainly from some  older experiments
performed with the Gargamelle ~\cite{eichten,erriquez} and the SKAT ~\cite{brunner} bubble
chambers filled with heavy liquid like Freon and/or 
Propane ~\cite{statement}.
 The number of
observed events was small leading to cross sections with large error bars due to poor
statistics. However, the results for the cross sections were found to be consistent with
predictions of the Cabibbo theory with SU(3) symmetry. A  suppression of cross sections 
due to nuclear medium effects is clearly seen, specially in the experiments of Erriquez et
al.~\cite{erriquez} but no attempts have been made to theoretically estimate the nuclear
medium effects on the weak production of hyperons from nuclei. An understanding of these
nuclear effects would  be useful for the analysis of future experiments which are being planned 
to study the weak production of strange particles in the context of neutrino oscillation and
proton decay search experiments. Such experiments are planned with the NUMI beamline
in the MINERVA experiment~\cite{minerva}. These reactions may also be seen at K2K and 
MiniBooNE where the effective reach of neutrino energy for cross section measurement 
could reach about 3 GeV~\cite{Ahn:2002up,Monroe:2004su}. The study of weak production 
of strange particles is an important subject in itself as it helps to experimentally 
determine the momentum dependence of various  transition form factors and test the 
theoretical models proposed for SU(3) breaking in semileptonic $\Delta S=1$ processes.

In this paper we report on the study of antineutrino induced quasi-elastic production of
$\Lambda$ and $\Sigma$ hyperons from nucleons i.e. reactions (1) to (3) and also the
effects of nuclear medium and final state interactions  when these reactions take place on
nucleons bound in nuclei. In section~\ref{formalism}, we describe the general formalism for 
calculating
the differential and total cross section for the process ${\bar{\nu_l}} +N\rightarrow l^+
+ Y$ using  Cabibbo theory with SU(3) symmetry where the transition form factors for
$N\rightarrow Y$  transitions are determined from a theoretical analysis of the latest
experiments on semileptonic decay of hyperons, i.e $Y\rightarrow N + l^- +{\bar{\nu_l}}$.
In section~\ref{nuclear}, we describe the nuclear medium effects when these reactions take
 place in
nuclei like  $^{16}{O}$ or $^{56}{Fe}$ which are target nuclei for future
detectors planned to be used in neutrino oscillation and proton decay search experiments.
In section \ref{results}, we present  the numerical results for total and differential cross 
sections for production of leptons and hadrons from nucleon and nuclear targets.
We also consider the pion production due to the weak decay of the hyperons. Finally we  
summarize and give main conclusions of our work in the last section.

\section{Formalism}
\label{formalism}
\subsection{Cross section  and Matrix elements}

The differential cross section $d\sigma$ for the process ${\bar{\nu_l}}(k) +N(p)\rightarrow
l^+(k^\prime) + Y(p^\prime)$, with $q=p^\prime-p=k-k^\prime$ is given by

\begin{equation}
\label{crosv.eq}
d\sigma=\frac{1}{(2\pi)^2}\frac{1}{4E_\nu \sqrt{s}}\delta^4(k+p-k^\prime-p^\prime)
\frac{d^3k^\prime}{2E_{k^\prime}}\frac{d^3p^\prime}{2E_{p^\prime}}|{\cal{M}}|^2
\end{equation}
leading to 
\begin{equation}
\frac{d\sigma}{dQ^2}=\frac{1}{64\pi s E_\nu ^2}|{\cal{M}}|^2
\end{equation}
where $s=(q+p)^2$, $E_\nu=\frac{s-M^2}{2\sqrt{s}}$ is the CM neutrino energy, 
$M$ is the nucleon mass and ${\cal{M}}$ is the scattering amplitude matrix 
element written as
\begin{equation}
{\cal{M}}=\frac{G}{\sqrt{2}}a_c {\bar{v}}(k^\prime)\gamma^\mu(1+\gamma^5)v(k)
<Y(p^\prime)|V_\mu-A_\mu|N(p)>,
\end{equation}
where $a_c=sin\theta_c$ for $\Delta S=1$ processes and  $a_c=cos\theta_c$ for $\Delta S=0$
processes. The matrix elements  $<Y(p^\prime)|V_\mu|N(p)>$ and $<Y(p^\prime)|A_\mu|N(p)>$ 
correspond to the  transition matrix elements of the vector and axial currents $V_\mu$ and 
$A_\mu$ which are defined as
\begin{eqnarray}
\label{defff}
<Y(p^\prime)|V_\mu|N(p)>={\bar{u}_Y}(p^\prime)\left[\gamma_\mu f_1(q^2)+i\sigma_{\mu\nu}
\frac{q^\nu}{M+M_Y}f_2(q^2)+\frac{f_3(q^2)}{M_Y}q_\mu\right]u_N(p)\\
\label{defff2}
<Y(p^\prime)|A_\mu|N(p)>={\bar{u}_Y}(p^\prime)\left[\gamma_\mu g_1(q^2)+i\sigma_{\mu\nu}
\frac{q^\nu}{M+M_Y}g_2(q^2)+\frac{g_3(q^2)}{M_Y}q_\mu\right]\gamma^5\, u_N(p)
\end{eqnarray}
where $f_i(q^2)$, and $g_i(q^2)$, ($i=1,2,3$) are the vector and axial vector transition form
factors. In defining these matrix elements, we follow the Bjorken Drell~\cite{bjork} 
conventions for 
the  Dirac matrices. 
 The determination of these form factors is done using
Cabibbo theory with SU(3) symmetry which describes  the recent precision data on
semileptonic decays of hyperons~\cite{Gaillard:1984ny,Cabibbo:2003cu} quite well. 
The corrections due to SU(3) breaking effects on semileptonic decays have been 
discussed in literature and  are found to be small~\cite{dono}.

In the following, we briefly outline the procedure for determination of various vector and axial 
vector transition form factors  $f_i(q^2)$ and $g_i(q^2)$ defined in equations~\ref{defff} and
~\ref{defff2}.

\subsection{Form Factors}
In the standard model, the vector and axial vector currents $V_\mu$ and $A_\mu$ are defined as
\begin{eqnarray}
V^i_\mu&=&\bar{q}\frac{\lambda^i}{2}\gamma_\mu q\\
A^i_\mu&=&\bar{q}\frac{\lambda^i}{2}\gamma_\mu\gamma_5 q
\end{eqnarray}
where $\frac{\lambda^i}{2}$ are the generators of flavour SU(3). Assuming that, $V^i_\mu$ and
$A^i_\mu$ belong to the octet representation of flavour SU(3), and neglecting any SU(3)
breaking effects, vector and axial vector  transition form factors for all the  N$\rightarrow$
Y transitions  can be expressed in terms of two functions for vector(axial vector) current which
could be determined from the experimental data on semileptonic decays of nucleons and
hyperons. This is because, the coupling of initial and final baryon states belonging to an
octet representation of SU(3), through an octet of vector ( axial vector ) currents is
described in terms of two reduced matrix elements F and D corresponding to the 
antisymmetric and
symmetric coupling of two octets of baryons in the initial and final state to the octet of vector 
(axial vector) currents, through
SU(3) Clebsch Gordan coefficients. More precisely, the vector and axial vector form factors
$f_i(q^2)$ and $g_i(q^2)$ defined above are given in terms of the functions $F^V_i(q^2)$ and
$D^V_i(q^2)$ corresponding to vector couplings and $F^A_i(q^2)$ and $D^A_i(q^2)$
corresponding to axial vector couplings as
\begin{eqnarray}
\label{ff0}
f_i(q^2)=aF^V_i(q^2)+bD^V_i(q^2), \text{(i=1,2,3)}\\
\label{ff01}
g_i(q^2)=aF^A_i(q^2)+bD^A_i(q^2), \text{(i=1,2,3)}
\end{eqnarray}
The constants $a$ and $b$ are the SU(3) Clebsch Gordan coefficients given in
Table~\ref{tabI} for the reactions of our present interest. We see that all the form factors for
$p\rightarrow \Sigma^0$ are $\frac{1}{\sqrt{2}}$ times the form factors
$n\rightarrow\Sigma^-$ transitions, leading to the prediction that
$\frac{d\sigma}{dq^2}({\bar{\nu}}+n\rightarrow
\mu^++\Sigma^-)/\frac{d\sigma}{dq^2}({\bar{\nu}}+p\rightarrow \mu^++\Sigma^0)=\frac{1}{2}$.
This  is reflection of the  $\Delta$I=$\frac{1}{2}$ rule, inherent in the Cabibbo theory of
$\Delta S=1$ weak processes.
\begin{table*}[h]
\begin{tabular}{|c|c|c|}\hline
Transitions &a &b \\ \hline
$p\rightarrow n$& 1& 1 \\
$p\rightarrow \Lambda$&  $-\sqrt{\frac{3}{2}}$&$-\sqrt{\frac{1}{6}}$\\
$n\rightarrow \Sigma^-$& -1& 1 \\
$p\rightarrow \Sigma^0$& $-\frac{1}{\sqrt{2}}$& $\frac{1}{\sqrt{2}}$\\ \hline
\end{tabular}
\caption{Values of the Form Factors coefficients $a,\, b$ of Eqs.~\ref{ff0}-\ref{ff01}.}
\label{tabI}
\end{table*}

Furthermore, the assumption that $V_\mu$ and $A_\mu$ belong to the octet representation of
flavour SU(3), implies that the symmetry properties of the $\Delta$S=0 currents which are
well verified in the study of $n\rightarrow p+e^-+{\bar{\nu_e}}$ decays are also obeyed by
the  the $\Delta S=\pm 1$ currents. Accordingly, we assume

(a) G invariance and SU(3) symmetry leading to prediction that  $f_3(q^2)\ =\ g_2(q^2) =0$.

(b) Conserved Vector Current and SU(3) symmetry leading to   $f_3(q^2) =0$ and determination
of other vector transition form factors in terms of the electromagnetic form factors of
protons and neutrons. The electromagnetic form factors of protons and neutrons in terms of
nucleons ($N = p,n$) are defined through the matrix element of the electromagnetic current
$V_\mu$ taken   between the nucleon states ($N = p,n$) as  $<N(p^\prime)|V^{em}_\mu|N(p)>$ and is
written as

\begin{equation}
\label{vem}
<N(p^\prime)|V^{em}_\mu|N(p)>=
{\bar{u}}(p^\prime)\left[\gamma_\mu f^{N=p,n}_1(q^2)+i
\sigma_{\mu\nu}\frac{q^\nu}{2M}f^{N=p,n}_2(q^2)\right]u(p)
\end{equation}
where $ f^{N=p,n}_1(q^2)$ are the electromagnetic form factors for nucleons.
$ V^{em}_\mu$ is the electromagnetic current given by

\begin{equation}
\label{vem2}
V^{em}_\mu=V^3_\mu+\frac{1}{\sqrt{3}}V^8_\mu
\end{equation}
 where  the superscripts 3 and 8 show  SU(3) indices. 
Evaluating Eqn. \ref{vem} between the nucleon states using their SU(3) indices we get
\begin{eqnarray}
\label{vem3}
f^n_i(q^2)&=&-\frac{2}{3}D^V_i(q^2), \;\;\;\;\ \text{i=1,2}\nonumber\\
f^p_i(q^2)&=&F^V_i(q^2)+\frac{1}{3}D^V_i(q^2), \;\;\;\ \text{i=1,2}
\end{eqnarray}
Eqns. \ref{vem3}, determine $F^V_i(q^2)$ and $D^V_i(q^2)$ in terms of the electromagnetic form 
factors for neutrons and protons $f^n_i(q^2)$ and $f^p_i(q^2)$ as
\begin{eqnarray}
F^V_i(q^2)&=&f^p_i(q^2)+\frac{1}{2}f^n_i(q^2)\nonumber\\
D^V_i(q^2)&=&-\frac{3}{2}f^n_i(q^2)
\end{eqnarray}
Once $F^V_i(q^2)$ and $D^V_i(q^2)$ are determined, the transition vector form factors $f_1(q^2)$ and 
$f_2(q^2)$ defined in Eqn. \ref{defff} are determined for all transitions, in terms of 
$f^{p,n}_i(q^2)$ and are presented in table \ref{tabII}. For $f^{p,n}_i(q^2)$ we 
take~\cite{Galster:1971kv,Alberico:2001sd}:
\begin{eqnarray*}
 f^{p,n}_{1}(q^2)&=&\frac{1}{(1-\frac{q^2}{4M^2})}\left[{G^{p,n}_E(q^2)-
 \frac{q^2}{4M^2}G^{p,n}_M(q^2)}\right]\\
 f^{p,n}_2(q^2)&=&\frac{1}{(1-\frac{q^2}{4M^2})}[{G^{p,n}_M(q^2)-G^{p,n}_E(q^2)}]
\end{eqnarray*}
where

\begin{equation}
G^p_E(q^2)=\left(1-\frac{q^2}{M^2_V}\right)^{-2}
\end{equation}
\[G^p_M(q^2)=(1+\mu_p)G^p_E(q^2),~G^n_M(q^2)=\mu_n G^p_E(q^2);  \]
\[G^n_E(q^2)=(\frac{q^2}{4M^2})\mu_n G^p_E(q^2) \xi_n;~\xi_n=
\frac{1}{1- \lambda_n\frac{q^2}{4M^2}}\]
\[\mu_p=1.792847, \mu_n=-1.913043, M_V=0.84GeV, ~\mbox{and}  ~\lambda_n=5.6.\]
The numerical value of the vector dipole mass $M_V$= 0.84 GeV is taken from experimental 
data on electron proton scattering. However, in the $\Delta S=1$ sector with  SU(3) 
symmetry a scaled
value of $M_V$= 0.97 GeV has also been used in the analysis of semileptonic 
decays~\cite{Gaillard:1984ny}.

(c) The Partial Conservation of Axial Current (PCAC) hypothesis and SU(3) symmetry leads to
the determination of the pseudo vector transition form factor  $g_3(q^2)$ in terms of the
axial vector form factor  $g_1(q^2)$  which predicts $g_3(q^2)=\frac{2M^2}{m^2_\pi-
q^2}g_1(q^2)$. These form factors are  determined from the  experimental data on $\Delta$S=0
neutrino scattering on nucleon and semileptonic hyperon decays. In these processes, the contribution
of $g_3(q^2)$, being proportional to $\frac{m_l}{M}$, is small and  is generally neglected in
the analysis of neutrino scattering and semileptonic decays. Therefore, the $q^2$ dependence of 
$g_3(q^2)$ specially at higher $q^2$ is not  determined experimentally. Some experimental
information on $g_3(q^2)$ is available from studies on  muon capture in nucleon and nuclei,
which is consistent with the predictions of PCAC .  However, the numerical contribution of
$g_3(q^2)$ to the cross sections in the present reactions is also  small and is neglected.
 With these assumptions the
only undetermined  form factor needed for the calculation of the matrix element defined in
equations \ref{defff} and \ref{defff2} is  $g_1(q^2)$.

In order to determine $q^2$ dependence of transition form factors $g_1(q^2)$ for all transitions
under present consideration one needs the  $q^2$ dependence of $ F^A_1(q^2)$ and $D^A_1(q^2)$
separately which is not available due to lack of high $q^2$ data from semileptonic processes in
the$\Delta S=1$ sector. We therefore, assume that $F^A_1(q^2)$ and $D^A_1(q^2)$ have the same $q^2$
dependence.  From table~\ref{tabI} the axial vector form  factor $g_1(q^2)$ is given by
$g_1(q^2)=F^A_1(q^2)+D^A_1(q^2)$ for the $\nu_\mu+n\rightarrow\mu^-+p$ reaction. The determination
of $q^2$ dependence of the axial vector form factor in $\nu_\mu+n\rightarrow\mu^-+p$ reaction yields
information about the $q^2$ dependence of $F^A_1(q^2)+D^A_1(q^2)$. 

 We now assume that
$F^A_1(q^2)$ and $D^A_1(q^2)$ separately have the  $q^2$ dependence which is given by the the $q^2$
dependence of $g^{n\rightarrow p}_A(q^2)$, i.e. $g^{n\rightarrow p}_1(q^2)=g^{n\rightarrow
p}_1(0)\left(1-\frac{q^2}{M^2_A}\right)^{-2}$. 
We thus take
$$ F^A_1(q^2)  = F\left(1-\frac{q^2}{M^2_A}\right)^{-2},\;\ \text{with}\; F=F^A_1(0)\;\ $$
and
$$ D^A_1(q^2) = D\left(1-\frac{q^2}{M^2_A}\right)^{-2},\; \text{with}\;\;\ D=D^A_1(0).$$
The numerical value of the axial vector dipole
mass $M_A$ is taken from the analysis  of world data on quasielastic neutrino nucleon
scattering to be 1.03 GeV~\cite{Ahrens:1986xe,Alberico:2001sd}. However, the recent 
high statistics K2K  
experiment on quasielastic scattering  at low energies suggests a higher value of
$M_A=1.20\pm0.12$ GeV~\cite{k2k}. On the other hand, the analysis of very low $q^2$ data on 
semileptonic decays of hyperons uses an axial dipole mass of $M_A$=1.25 GeV in $\Delta$S=1
sector~\cite{Gaillard:1984ny}.

With this parametrization of $F^A_1(q^2)$ and $D^A_1(q^2)$, the constants F and D are determined 
from the analysis of present experimental data on
semileptonic decays of nucleons and hyperons corresponding to very low $q^2$ which gives 
$F+D = 1.2670\pm 0.0030$ and
$F-D = -0.341\pm 0.016$~\cite{Cabibbo:2003cu}. Using these values of  $F^A_1(q^2)$ and 
$D^A_1(q^2)$, we present in 
table~\ref{tabII}, the values of
$g_1(q^2)$ for various transitions of our present interest in terms of
$x=\frac{F^A_1(q^2)}{F^A_1(q^2)+D^A_1(q^2)}=\frac{F}{F+D}$ and
$g_A(q^2)=(F+D)\left(1-\frac{q^2}{M^2_A}\right)^{-2}$.
\begin{table*}[h]
\begin{tabular}{|c|c|c|c|}\hline
Transitions &$f_1(q^2)$&$f_2(q^2)$&$g_1(q^2)$ \\ \hline
$n\rightarrow p$ & $f^p_1(q^2)-f^n_1(q^2)$&$f^p_2(q^2)-f^n_2(q^2)$&$g_A(q^2)$ \\  
$p\rightarrow \Lambda$& $-\sqrt{\frac{3}{2}}f^p_1(q^2)$&$-\sqrt{\frac{3}{2}}f^p_2(q^2)$&
                                          $-\sqrt{\frac{3}{2}}\frac{(1+2x)}{3}g_A(q^2)$\\
$n\rightarrow \Sigma^-$&-($f^p_1(q^2)+2f^n_1(q^2)$)&-($f^p_2(q^2)+2f^n_2(q^2)$)&
                                           $(1-2x)g_A(q^2)$ \\ \hline
\end{tabular}
\caption{Form Factors of Eqs.~\ref{defff}-\ref{defff2}.}
\label{tabII}
\end{table*}

\section{Nuclear Medium and Final state Interactions}
\label{nuclear}
\subsection{Nuclear Effects} 
When the reactions shown in equations (1-3) take place on nucleons which are bound in the
nucleus, certain constraints on their dynamics arising due to  the Fermi motion and Pauli blocking
effects of initial nucleons have to be considered. In the final state the produced hyperons
are not subjected to any Pauli Blocking but are affected by the final state interactions with
the nucleus through the hyperon nucleon quasi-elastic and
charge exchange scattering processes. Moreover, the charged lepton in the final state moves in
the Coulomb field of the final nucleus. However, in the energy region of 1- 3 GeV, the
effect of Coulomb distortion of the charged lepton wave function is small and is neglected in 
the present calculations. The  Fermi 
motion effects are calculated in a local Fermi Gas model where the the
differential cross section for the process ${\bar{\nu_l}} +N\rightarrow l^+ + Y$ is
now written as

\begin{equation}
\label{eq.nuc}
d\sigma=\frac{1}{(2\pi)^2}2\int{d^3\vec{r}}\frac{d^3\vec{p}}{(2 \pi)^3}
n(p,r)\delta^4(k+p-k^\prime-p^\prime)
\frac{d^3\vec{k}\,^\prime}{2E_{k^\prime}}\frac{d^3\vec{p}\,^\prime}{2E_{p^\prime}}
\frac{1}{4 E_\nu^{CM} \sqrt{s}}
|{\cal{M}}|^2
\end{equation}
where $n(p,r)$ is the local occupation number of the initial nucleon of momentum $p$ 
localized at a radius $r$ in the nucleus, and is determined in the local density 
approximation. Here,
$E_\nu^{CM}$ and $s$ are the neutrino energy in the nucleon-neutrino CM system and
the nucleon neutrino invariant mass squared respectively.
Solving the $\delta$ function of momentum conservation , we do the integration over the 
hyperon momentum
$\vec{p}\,^\prime$, and we use the $\delta$ function of energies to integrate the  cosinus 
of the angle of the initial nucleon momentum $\vec{p}$. Then, the differential cross 
section for the quasielastic hyperon production  from nuclei can be written as
\begin{equation}
\label{nuc.eq}
d\sigma=\frac{1}{64\pi^4}\int{r^2dr}d\phi_p\int_0^{k_F(r)}dp\;d^3\vec{k}\,^\prime
\frac{p}{E_\nu^{CM}\sqrt{s}E_\mu|\vec{k}-\vec{k}\,^\prime|}|{\cal{M}}|^2
\end{equation}
with $k_F(r) = (\frac{3}{2} \pi^2\rho(r))^{\frac{1}{3}}$,
where $\rho(r)$ is the target nucleon density in the nucleus which is taken from 
ref.~\cite{DeJager:1987qc} for the protons, and scaled with a factor $N/Z$ for the neutrons.
 All kinematic variables are defined by the integral itself, 
except the cosinus of relative angle between $\vec{p}$ and $\vec{k}-\vec{k}\,^\prime$ which 
is obtained from the $\delta$ function of energies.

To obtain these formulas we have followed a quasi-free approach where
both $\Sigma$ and $\Lambda$ have been treated as stable particles, with a well defined
energy for a given momentum. This is acceptable because both are quite narrow even in the
nuclear medium, see i.e. ref.~\cite{Oset:1989ey}, where additional decay channels are 
present. Also, in the actual implementation of Eq.~\ref{nuc.eq} when solving the $\delta$ 
of energies, we have neglected the real part of the hyperon optical potential in the nucleus.
We have checked numerically that potentials of a typical size ($\approx -30 MeV\rho/\rho_0$)
do not modify appreciably the results.

\subsection{Final State Interactions}

The hyperons $\Lambda^0$, $\Sigma^0$, $\Sigma^-$ which are produced in reactions (1-3)
undergo elastic and charge exchange scattering with the nucleons present in the nucleus
through strong interactions while some of the $\Sigma^0$ disappear through the
electromagnetic decay channel  $\Sigma^0\rightarrow \Lambda^0 + \gamma$. Therefore the
production cross sections for the hyperons from the nuclear targets are affected by the
presence of the electromagnetic and strong  interactions of final state hyperons in the
nuclear medium. One of the  interesting features of the final state interactions(FSI) of hyperons 
in the
nuclear medium is the appearance of $\Sigma^+$ hyperons which are not produced in the basic
weak process induced by the $\bar\nu$. This is due to  charge exchange scattering
processes like $\Lambda^0+p\rightarrow \Sigma^+ +n$ and $\Sigma^0+p\rightarrow \Sigma^+ +n$ 
which can take place in nuclei. The effect of FSI on the weak production
cross section for $\Sigma^0$, $\Sigma^-$ and $\Lambda^0$ and the appearance of $\Sigma^+$ are
estimated  with the help of a Monte Carlo code for propagation of hyperons in the nuclear
medium using as input the scarce available experimental cross sections for the hyperon nucleon 
scattering  cross sections. We have compiled the parametrizations used in this work in the 
Appendix.

\subsection{Monte Carlo simulation}
\label{MC}

>From Eq.~\ref{eq.nuc} we can obtain $\frac{d^6\sigma}{d^3r\,d^3 k^\prime}$ after
performing the integration over the rest of variables. This profile function is then
used as input for our Monte Carlo simulation.  We generate  hyperon production events
by selecting a random position $r$ and a momentum $k^\prime$ and assigning to the
event the weight given by the profile function.  We then assume the real part of the
hyperons nuclear potential to be weak compared with their kinetic energies and
propagate them following straight lines till they are out of the nucleus. To take into
account the collisions we follow the hyperon by moving it a short distance $dl$, along
its momentum direction, such that  $P\, dl <<\, 1$, where $P$ is the probability  of
interaction per unit length. A random number $x\in[0,1]$ is generated and we consider that
an interaction has taken place when  $P\, dl > x$. If no interaction occurs we repeat
the procedure by moving the hyperon a new step $dl$.

The  probability of interaction  per unit length of a hyperon $Y$ is given by 
\begin{equation}
P_Y=\sum_f \left\{ \sigma_{Y+n\rightarrow f}(\bar{E})\rho_n +
\sigma_{Y+p\rightarrow f}(\bar {E})\rho_p\right\}
\end{equation}
where $f$ accounts for all possible final channels, $n$ and $p$ are neutrons and
protons and $\rho_n$, $\rho_p$ are their local densities. The cross section is 
evaluated  at an invariant energy of the neutrino-nucleon system averaged over the 
local Fermi sea. We use a threshold energy cut of 30 MeV for quasielastic collisions 
($\Lambda\rightarrow\Lambda$,$\Sigma\rightarrow\Sigma$). Below this energy, we 
only consider possible $\Sigma\rightarrow\Lambda$ processes. 
Thus, the energy spectra at those low kinetic energies will not be meaningful.

If the hyperon has interacted we select the channel accordingly to their respective
probabilities. Finally, once the channel has been selected, we approximately implement 
Pauli blocking with the following procedure. A random nucleon is selected in the local
Fermi sea.  Assuming isotropic cross sections in the hyperon-nucleon CM system, 
we generate a random  
scattering angle in that system and calculate the hyperon and nucleon  momenta.
Finally, we boost  these momenta to the lab system. If the final nucleon is below the 
Fermi level
(Pauli blocked) we consider that there was no interaction and the hyperon continues its
movement.  Otherwise, we have a new 
hyperon type and/or a new direction and energy.

It should be mentioned that all this procedure does not modify neither the
$(\bar{\nu},lepton)$ cross section, nor the $q^2$ dependence of that observable, and only 
the type of outgoing hyperon and its energy and angle distributions are modified. 
In exclusive 
reactions, where both the lepton and the hyperon are observed, there could be some changes
due to the fact that the lepton distributions would correspond to those of the primary 
hyperon and not to that of the observed one that could be of a different kind.  

\section{Results and Discussion}
\label{results}

The numerical evaluations of the quasielastic production of   $\Sigma^0$, $\Sigma^-$ and
$\Lambda^0$ hyperons induced by antineutrinos from free nucleons have been done using
Eq.~\ref{crosv.eq} with the form factors given in table~\ref{tabII}. The nuclear medium 
effects due to Fermi motion are incorporated through Eq.~\ref{nuc.eq}. The FSI
effects, due to hyperon nucleon elastic and charge exchange scattering processes in presence
of other nucleons in nuclei are taken into account using a Monte Carlo simulation described
in section~\ref{MC}. All the results presented here correspond to muonic antineutrinos.

\subsection{Lepton differential cross sections}
We first present the differential cross section for antineutrino induced $\Delta S=1$ weak 
quasielastic processes from nucleon and nuclear targets. The sensitivity of the 
differential cross sections to the axial vector dipole mass has been studied. We have 
also studied the effect of nuclear medium and final state interactions on the differential 
cross sections. We find  that in the range 
of energies under analysis,  Fermi motion of the nucleons and FSI of the 
hyperons do not appreciably modify  the lepton distributions, except for a
scale factor that can also be seen in the total cross sections.
\begin{figure}
\includegraphics[width=9cm,angle=-90]{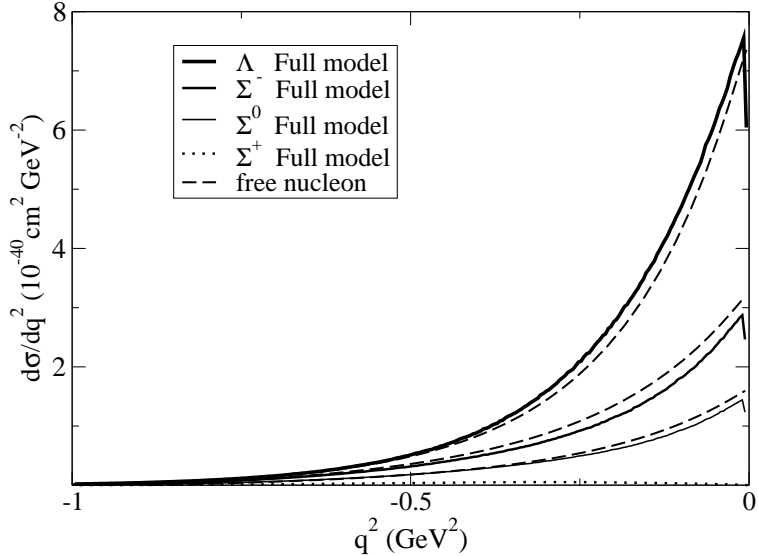}
\caption{$q^2$ distributions  for the reaction $\bar{\nu}+A\rightarrow \mu^++Y+X$ at
$E_{\bar{\nu}}=1$ GeV in nucleons and in $^{16}O$. In the nuclear case, the cross sections
are divided by 8.
 Solid lines: Full model; dashed lines: hyperon production on a free nucleon. The 
  upper curves correspond to $\Lambda$, next  to $\Sigma^-$, next to 
 $\Sigma^0$. Dotted line: $\Sigma^+$.}
\label{difq2.fig}
\end{figure}
As a typical case, we show in Fig.~\ref{difq2.fig} the $q^2$ dependence on free nucleons and 
on $^{16}O$ at $E_{\bar{\nu}}=1$ GeV. The lowest curve corresponds to 
the small $\Sigma^+$ production which occurs via FSI. The other lines show the
results for the $\Lambda$, $\Sigma^-$ and  $\Sigma^0$.
 The results without FSI are very close to the free nucleon
ones and are not shown. Even the full model curves have the  same shape.
Thus, we find that nuclear data could still be used 
to investigate  the $q^2$ dependence of the form factors in the hyperons sector.
\begin{figure}
\includegraphics[width=9cm,angle=-90]{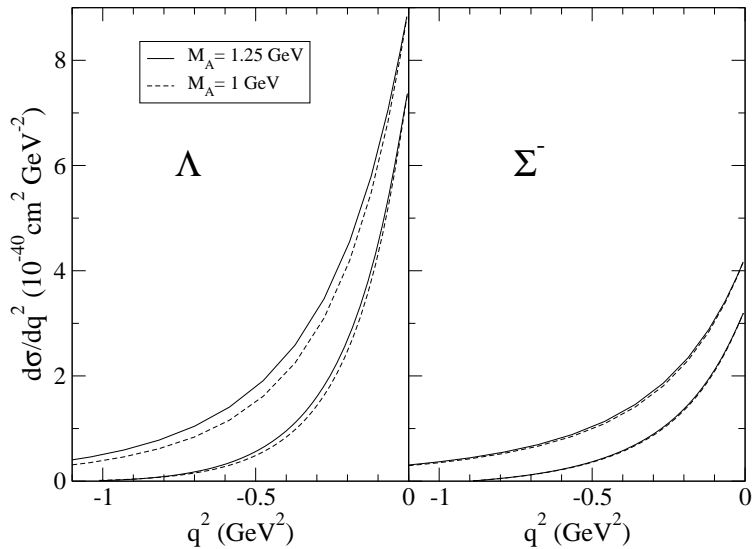}
\caption{$q^2$ distributions  for the reaction $\bar{\nu}+N\rightarrow \mu^++Y+N$ at
$E_{\bar{\nu}}=1$ (lower curves) and 3 GeV (upper curves) for two $M_A$ values.}
\label{dsdq2nuc.fig}
\end{figure}
However, as shown in Fig.~\ref{dsdq2nuc.fig}, the $M_A$ dependence is very mild. This is  
specially so  at low
energies and for the case of $\Sigma$ production. Only at relatively large antineutrino energies 
and for $\Lambda$ production the cross section shows some sensitivity to this parameter.

\subsection{Hyperons spectra}
\begin{figure}
\includegraphics[width=12.5cm,angle=-90]{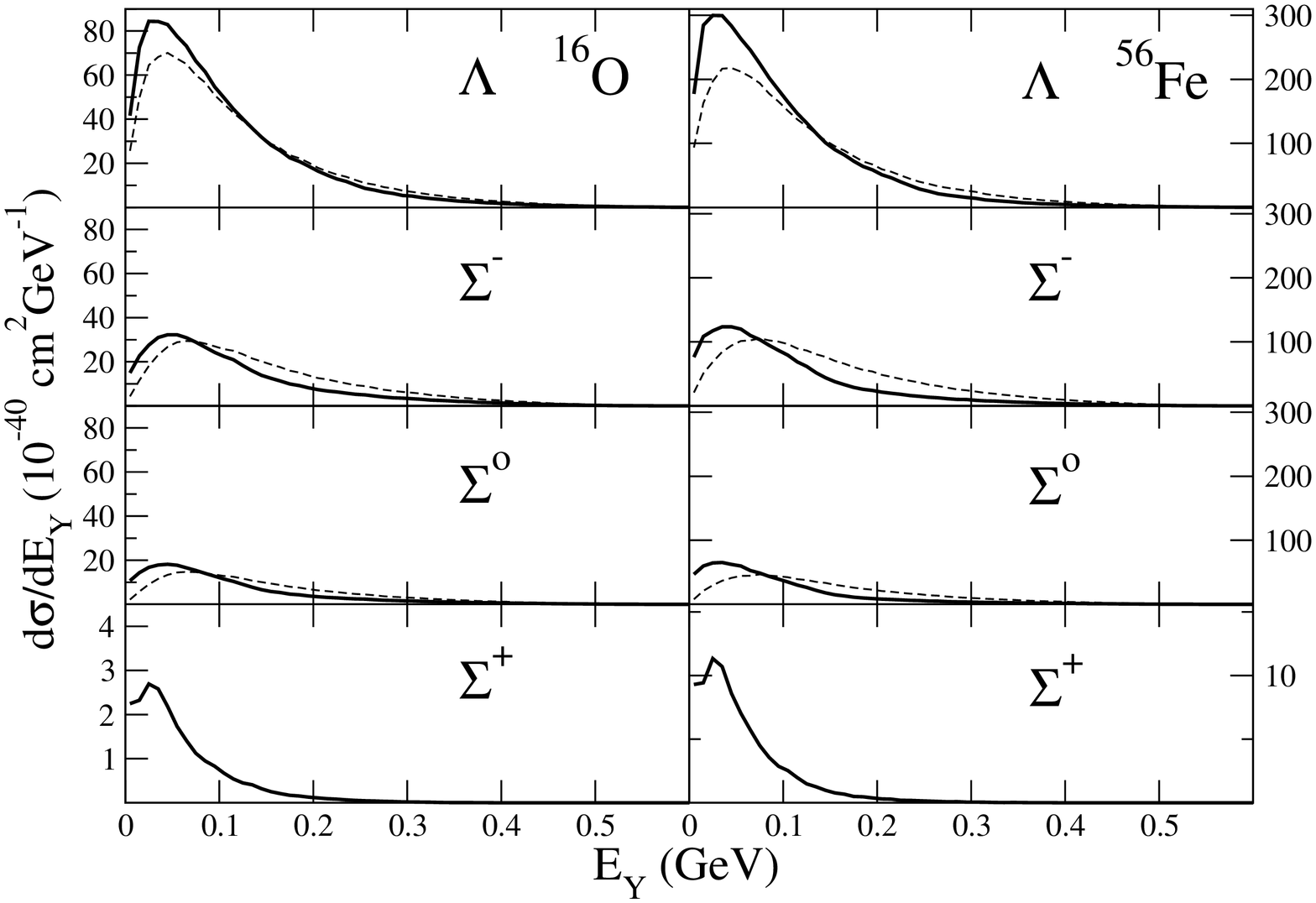}
\caption{Hyperons energy distributions as a function of the hyperon kinetic
energy for the reaction $\bar{\nu}+A\rightarrow \mu^++Y+X$ at
$E_{\bar{\nu}}=1$ GeV.
Left(right) side corresponds to $^{16}O$($^{56}Fe$). Solid line: full model,
dashed line: without final state interaction. }
\label{spectrum.fig}
\end{figure}

We show in Fig.~\ref{spectrum.fig}  the hyperons spectra with and without
FSI for 1 GeV antineutrinos.
The main effect of FSI is a redistribution of strength, pushing the spectra
towards lower energies. This is due to quasielastic collisions with the nucleons
and also to  inelastic scattering, in which the kind of hyperon changes and
part of the energy is passed to the nucleons. Also remarkable
is the appearance of $\Sigma^+$ through the
$\Sigma^0+p\rightarrow\Sigma^++n$ and
$\Lambda+p\rightarrow\Sigma^++n$ processes. This channel is not present on
free nucleons and will be further discussed in the  next section.
We should recall here that our MC code does not include neither the effects of the
real part of the optical potentials nor interactions of particles with kinetic 
energies below 30 MeV. Therefore, the results at those low energies are not meaningful
and are shown only for illustrative purposes.

\subsection{Total Cross sections} 

\begin{figure}
\includegraphics[width=9cm,angle=-90]{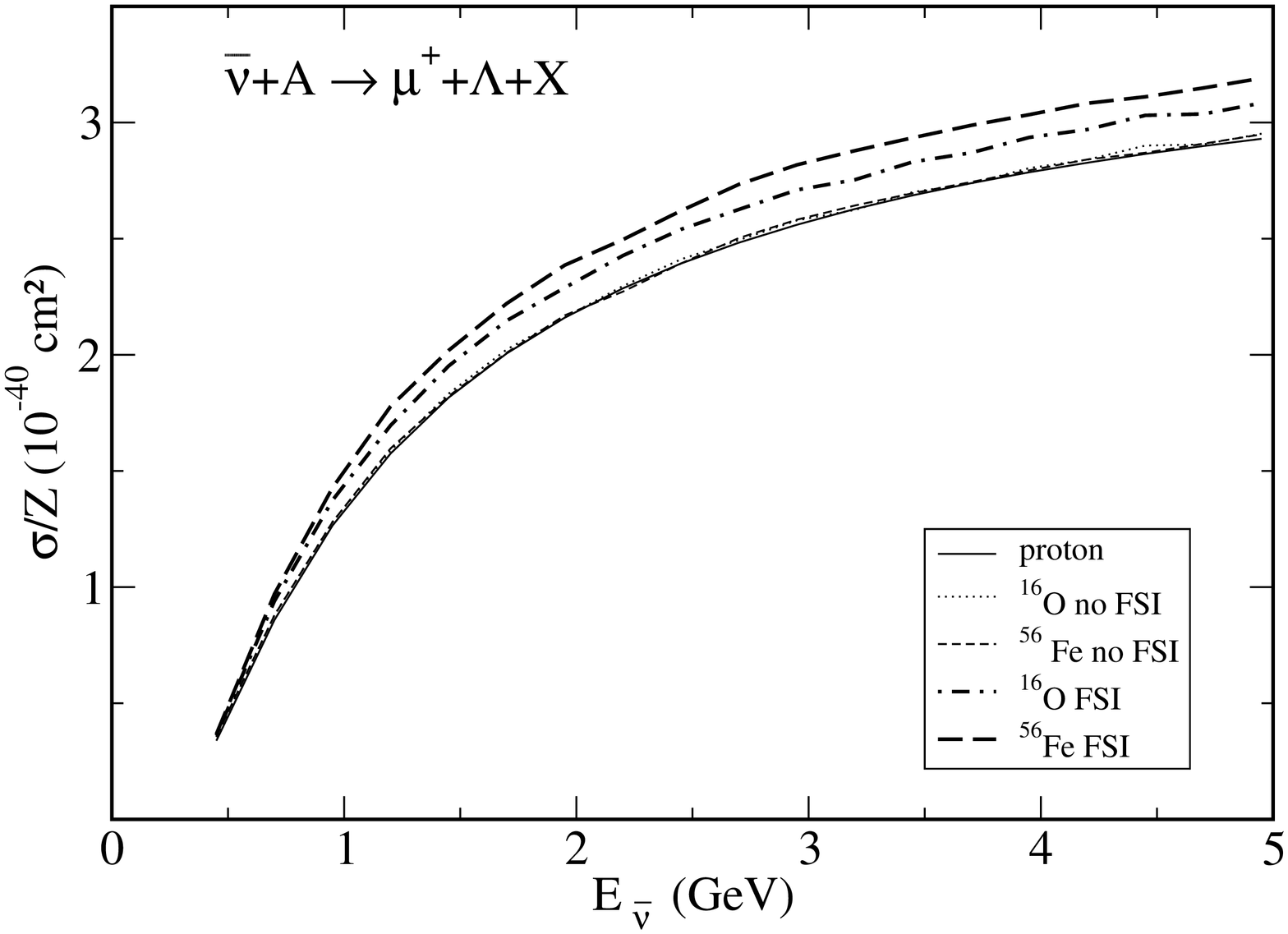}
\caption{Cross section for $\Lambda$ production induced by a muonic antineutrino 
divided by the number of protons.}
\label{Lambdamuon.fig}
\end{figure}

\begin{figure}
\includegraphics[width=9cm,angle=-90]{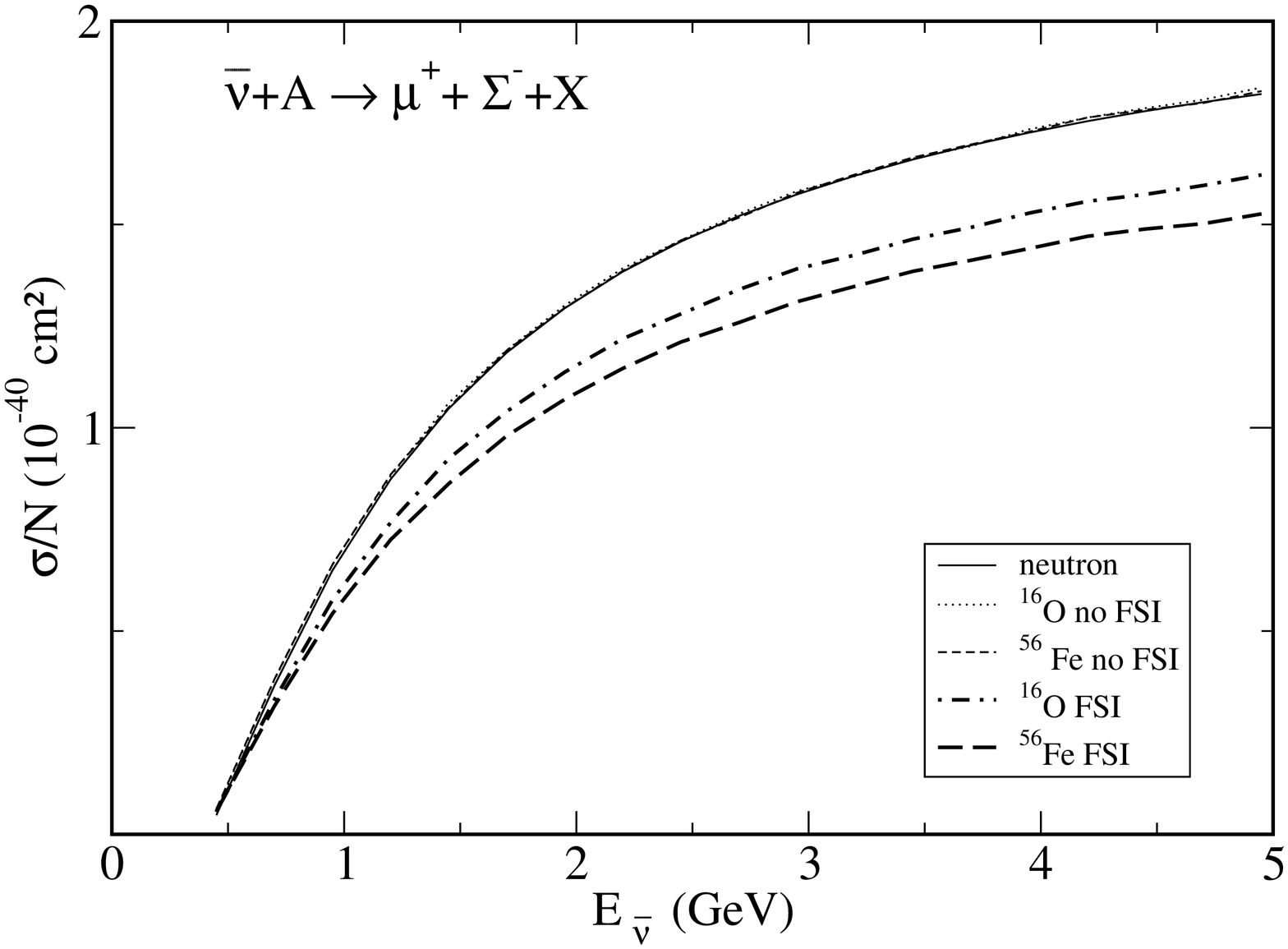}
\caption{Cross section for $\Sigma^-$ production induced by a muonic antineutrino 
divided by the number of neutrons.}
\label{Sigmammuon.fig}
\end{figure}

\begin{figure}
\includegraphics[width=9cm,angle=-90]{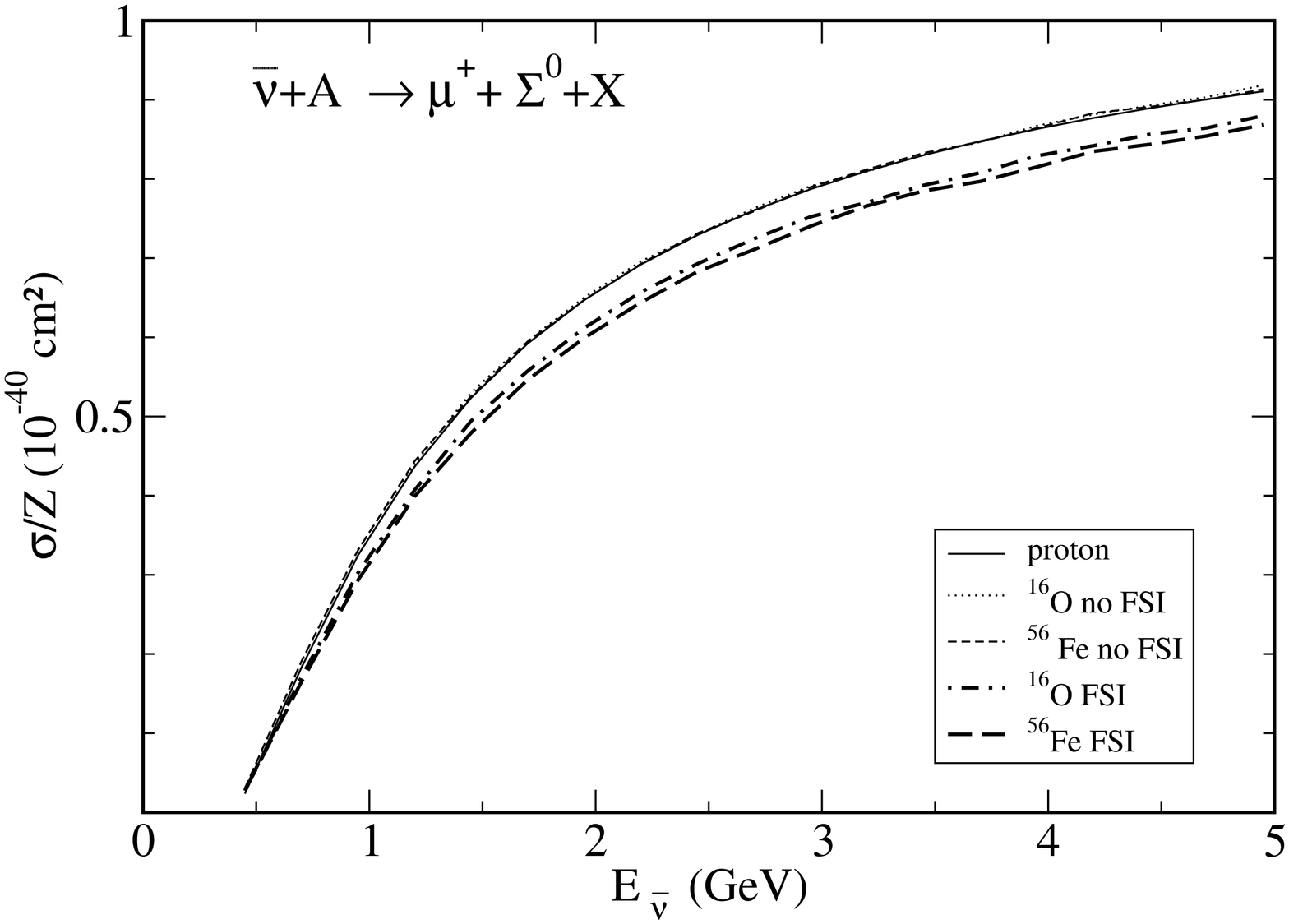}
\caption{Cross section for $\Sigma^0$ production induced by a muonic antineutrino 
divided by the number of protons.}
\label{Sigma0muon.fig}
\end{figure}

We present in figures~\ref{Lambdamuon.fig}-\ref{Sigma0muon.fig}
 the numerical results for the muonic antineutrino total cross sections
$\sigma(E_{\bar{\nu}})$  for free nucleons and for $^{16}O$ and $^{56}Fe$, divided 
in the nuclear case by the number of "active" nucleons, with and without the inclusion 
of FSI.
We see from these figures that

\begin{figure}
\includegraphics[width=10cm,angle=-90]{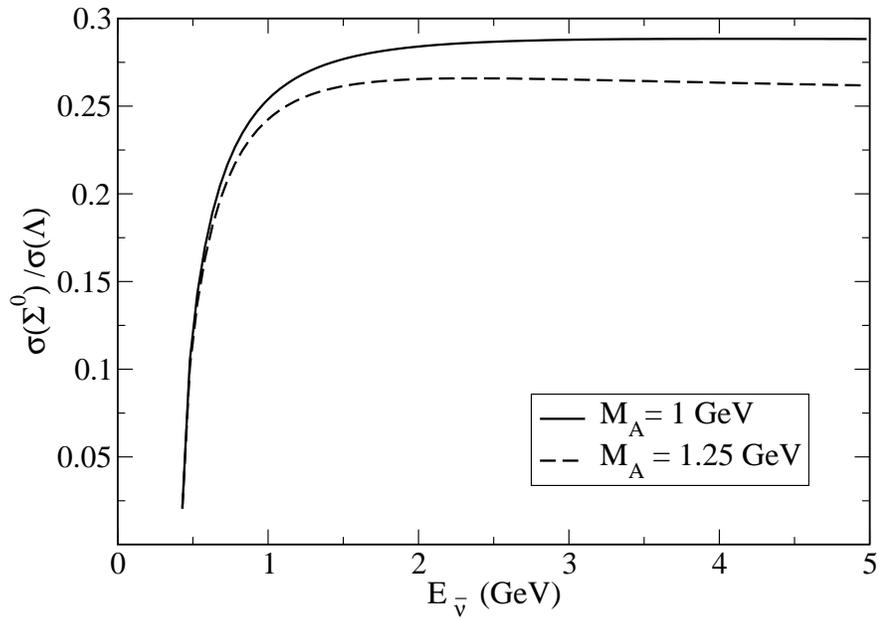}
\caption{The ratio
$R={\sigma(\bar\nu+p\rightarrow\mu^+ +\Sigma^0)}/{\sigma(\bar\nu+p\rightarrow\mu^+
+\Lambda)}$ as a function of the antineutrino energy.}
\label{ratio.fig}
\end{figure}

\begin{figure}
\includegraphics[width=10cm,angle=-90]{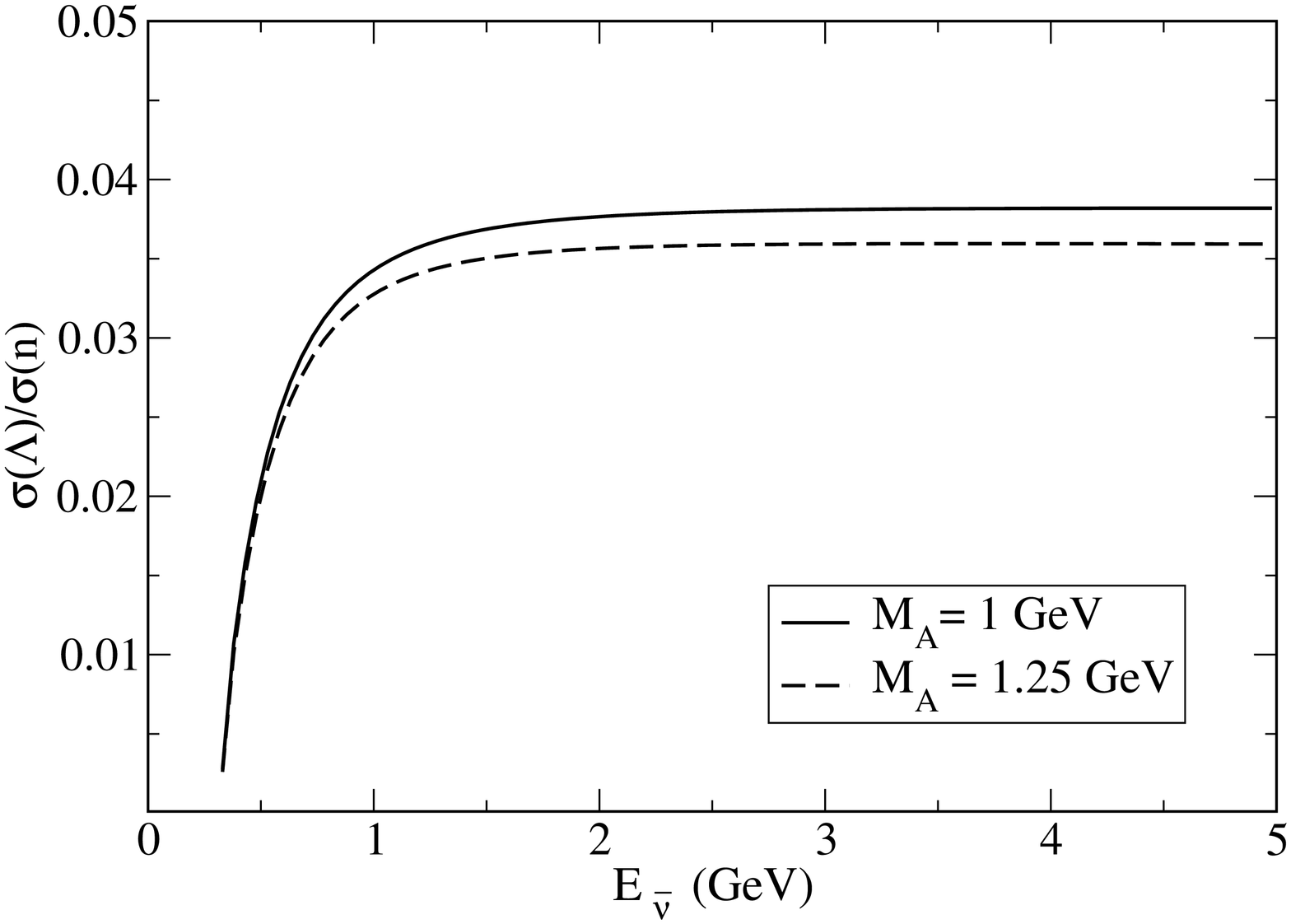}
\caption{The ratio
$R={\sigma(\bar\nu+p\rightarrow\mu^+ +\Lambda)}/{\sigma(\bar\nu+p\rightarrow\mu^+
+n)}$ as a function of the antineutrino energy.}
\label{ratio2.fig}
\end{figure}

(i) The effect of the Fermi motion of the initial nucleons is quite small on the 
quasielastic production of hyperons even for a heavy nucleus like  $^{56}Fe$ as shown in 
figures~\ref{Lambdamuon.fig}-\ref{Sigma0muon.fig}. Of course, this effect is larger
at energies, not shown in the figures, very close to threshold, where the cross sections 
are very small. Actually, in the nuclear case, the production threshold changes due to 
Fermi motion although the exact size of the effect depends on the hyperon nucleus optical 
potential.

(ii) The effect of hyperons FSI leads to an increase of the cross sections for
$\Lambda$ production and a decrease of $\Sigma^0$ and $\Sigma^-$ production cross sections.
This change in the cross section per nucleon increases  with the charge and mass number of
the nucleus and is larger for $^{56}Fe$ as compared to  $^{16}O$.
 This is because $\Sigma^{-,0}$ can disappear through the quasielastic processes
like $\Sigma^-+p\rightarrow\Lambda^0+n$, 
$\Sigma^0+n\rightarrow\Lambda^0+n$ and others, while the inverse process of
depletion of $\Lambda$ is also allowed, but inhibited due to the difference in masses. 
 In  addition to these strong processes leading to the depletion of
$\Sigma^0$, they are further depleted by  the electromagnetic decay
$\Sigma^0\rightarrow\Lambda+\gamma$. This has not been included in the calculation
as the mean life guarantees that the decay will occur out of the nucleus and can be easily 
taken into account when comparing with data.

(iii) For free nucleon targets, the cross section for production of $\Lambda$ is always
greater than the cross section for production of  $\Sigma^0$. The ratio
$R=\frac{\sigma(\bar\nu+p\rightarrow\mu^+ +\Sigma^0)}{\sigma(\bar\nu+p\rightarrow\mu^+
+\Lambda)}$ reaches  an asymptotic value of around 0.3 which is consistent with older
results of Cabibbo and Chilton~\cite{cab1} but is considerably different with the 
prediction of a relativistic quark model due to Finjord and Ravndal\cite{frinjord}. This
ratio is considerably smaller  at low energies due to threshold effects which suppress
$\Sigma^0$ production compared to  $\Lambda$ production. The  sensitivity of this ratio for
two values of the axial vector  dipole mass $M_A$ is shown in Fig.~\ref{ratio.fig}.

(iv) For free nucleon targets, using SU(3) symmetric form factors, the ratio of cross
sections  for $\Delta S=0$  and   $\Delta S=1$ induced processes by antineutrinos, i.e. 
$R=\frac{\sigma(\bar\nu+p\rightarrow\mu^+ +\Lambda)}{\sigma(\bar\nu+p\rightarrow\mu^+ +n)}$
reaches an asymptotic value of 0.04. This value comes  mainly due to the Cabibbo suppression
and  from the threshold effects which are quite large in this case. The energy dependence of
this ratio along with its sensitivity to the value of the axial vector dipole mass $M_A$ is shown
in Fig. \ref{ratio2.fig}.

\begin{figure}
\includegraphics[width=10cm,angle=-90]{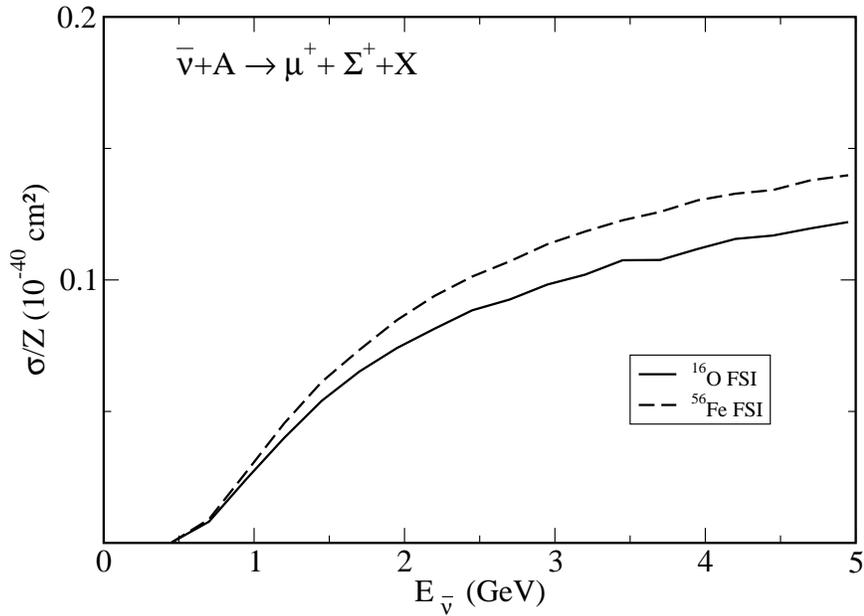}
\caption{Cross section for $\Sigma^+$ production induced by a muonic antineutrino 
divided by the number of protons as a function of the antineutrino energy.}
\label{sigmaplus.fig}
\end{figure}

(v) In Fig.~\ref{sigmaplus.fig}, we show the cross section for $\Sigma^+$ 
production.
Whereas in the other channels  FSI produces simply a correction to the direct process, 
in this case all events come from FSI and therefore the cross section is very sensitive
to the relatively unknown hyperon nucleon cross sections. This channel is 
a source of positive pions induced by a charged current antineutrino process, but the
cross section is very small and other sources, like charge exchange reactions
of pions produced inside the nuclei by other processes, as discussed below, will be 
more important.

\subsection{Pion production from hyperons }
 Currently, there is considerable interest in the weak pion production cross sections.
For these processes,  $\Delta$ excitation followed by its decay will be dominant at 
intermediate energies given its  strong coupling to the pion nucleon system. However,
two aspects  deplete its contribution to the pion production in nuclei.
First, the mean life of the  $\Delta$ is very short. Thus, it decays inside the
nucleus and part of the pions are absorbed and don't come out of the nucleus. 
 This is quite different to the hyperons case which decay 
weakly into pions. The hyperons large mean life implies that most of the times they decay 
already far from the nucleus avoiding the pion absorption. 
On the other hand, the mass of the $\Delta$  implies that the cross section decreases 
at low enough energies faster than for the $\Lambda$ and $\Sigma$ cases. These two factors
could partially compensate for the $tan^2\theta_c$ suppression.

\begin{figure}
\includegraphics[width=10cm,angle=-90]{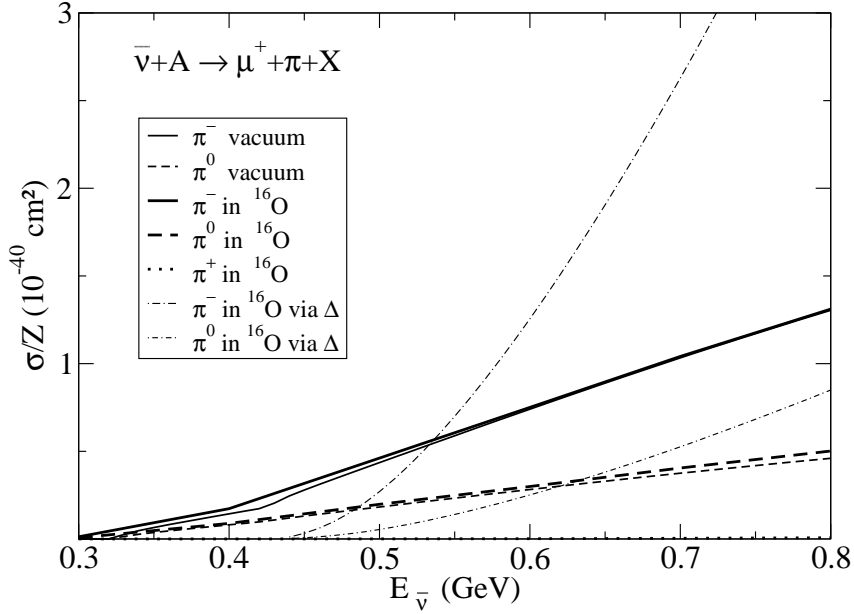}
\caption{Cross section for $\pi$ production via an intermediate hyperon
induced by a muonic antineutrino 
divided by the number of protons as a function of the antineutrino energy.
Results compared with pions produced via $\Delta$ excitation.}
\label{pion.fig}
\end{figure}

We show in Fig.~\ref{pion.fig} our results for pion production, obtained using the 
experimental branching ratios for the hyperons and the previous calculations for 
the hyperon production cross sections.
We also show results derived from the  $\Delta$ production cross section in $^{16}O$ of
ref.~\cite{Singh:1998ha} which incorporated pion absorption. 
In that paper, only the total number of pions (or $\Delta$'s) was obtained.
In order to compare with the current results, we have used the corresponding isospin 
factors to assign the charges of the pions (relative weights for
$p\rightarrow\Delta^0 \rightarrow p\pi^-$, $p\rightarrow\Delta^0 \rightarrow n\pi^0$ and
$n\rightarrow\Delta^- \rightarrow n\pi^-$ are 1/9, 2/9 and 1), thus neglecting possible
pion charge exchange reactions. 
We see that at low energies pions from hyperon decays dominate and the $\Delta$ mechanism
becomes dominant at energies  above 550 MeV for negative pions and 650 MeV for neutral pions.
The importance of the hyperon mechanisms would be larger for heavier nuclei, where 
pion absorption would suppress more strongly other competing mechanisms which produce the
pions inside the nucleons.

\section{Summary and Conclusions}
We have  studied the weak charged current induced quasielastic production of 
$\Lambda$ and $\Sigma$ hyperons from nucleons and nuclei. The transition form factors for 
the nucleon-hyperon transitions determined from an analysis experimental data on neutrino 
nucleon scattering and semileptonic decays of hyperons using Cabibbo theory with SU(3) 
symmetry have been applied to calculate the the total and differential cross sections for 
lepton and hyperon production from nucleon and nuclear targets. The nuclear medium and final 
state interaction effects have been calculated for the hyperon production from  nuclear targets 
like $^{16}O$ and $^{56}Fe$ which are proposed to be used in future detectors for neutrino 
oscillations and proton decay search experiments. These are calculated in a local Fermi gas 
model for the nuclei and a simple energy dependent parametrization for the hyperon nucleon 
scattering cross sections. The hyperon energy distribution for the quasielastic production 
of $\Lambda$, $\Sigma^+$ and $\Sigma^0$ hyperons induced by antineutrinos and the effect 
of final state interactions on their energy distribution has been studied. The energy 
distribution of $\Sigma^+$, which are  produced only as a consequence of final state 
interactions has also been presented. Finally the total cross sections for pion production 
due to decays of hyperons has been presented and compared with the pion production cross 
sections from $\Delta$ production. 
The main conclusions that can be drawn from our present study are:

(i) The differential cross sections $\frac{d\sigma}{dq^2}$ are more sensitive to the axial 
vector dipole mass for the case of $\Lambda$ production than $\Sigma$
production. However this sensitivity is not as large as compared to to the sensitivity 
of  $\frac{d\sigma}{dq^2}$ to the axial vector dipole mass for neutrino nucleon scattering 
in the $\Delta S=0$ sector.

(ii) The effect of nuclear medium effects on   $\frac{d\sigma}{dq^2}$ and total cross section 
$\sigma$ on the hyperon production is quite small.

(iii) The effect of final state interaction is to increase the cross sections for $\Lambda
$  production and to decrease the cross section for  $\Sigma^-$ and   $\Sigma^0$ production.
The  strength of production cross section shifts towards the lower energy of the produced
hyperon as  a result of final state interactions. The most interesting aspect of the final
state interaction  is that it leads to the production of  $\Sigma^+ $ hyperons which is of
the order of 10$\%$ of the  $\Sigma^- $  production cross sections from oxygen targets
around 1 GeV. This proportion increases with mass and charge of the nucleus.

(iv) The hyperon production is dominated by $\Lambda $ production and the production cross 
section for  $\Sigma^0$ is  small at lower energies but could approach 30$\%$ of  $\Lambda$ 
production as the energy increases and becomes larger than 1.0 GeV.

(v) At low energies, the nuclear pion production induced by antineutrinos through the
production of  hyperons and their subsequent  decays  can become  important as
compared to the antineutrino  pion production through the excitation and
subsequent decays of  $\Delta$ resonance. This,  for example, happens for
neutrino energies E $<$550(650) MeV) for the case of antineutrino  induced
$\pi^-(\pi^0)$ production at intermediate energies from $^{16}O$ target.

\begin{acknowledgments}
  This work was partially supported by DGI and FEDER funds,
contract  BFM2003-00856 and  by the EU Integrated Infrastructure
Initiative Hadron Physics Project contract RII3-CT-2004-506078. S. K. S 
acknowledges support 
from the Academic Exchange Agreement beetwen Aligarh M. U. and Valencia U.

\end{acknowledgments}

\appendix*
\section{ hyperon nucleon cross sections}
We present here the parametrizations used in our MC code for the
hyperon nucleon cross sections. In the formulas, cross sections are expressed 
in mb and
energies and momenta in GeV. The data used in the fits have been obtained 
from \cite{nnonline}, although we will also quote below the original
references. These parametrizations correspond to the best fits ($\chi$-square)
to data with the chosen functional form but the statistical errors of the data
are quite large and one should use these numbers as simple estimates. The
momenta in the formulas always refer to the hyperons

\begin{enumerate}
\item {$\Lambda+N\rightarrow \Lambda+N$} 

$\sigma =(39.66-100.45 x+92.44 x^2-21.40 x^3)/p_{LAB}$

where $x=Min(2.1,p_{LAB})$. Fitted to data for 
$\Lambda p\rightarrow \Lambda p$ scattering from 
refs.~\cite{lambdan,lambdansigma0}.

\item {$\Lambda+N\rightarrow \Sigma^0 N$}

$\sigma =(31.10-30.94x+8.16 x^2)p_{CM}^\Sigma/p_{CM}^\Lambda$

where $x=Min(2.1,p_{LAB})$. Fitted to data for 
$\Lambda p\rightarrow \Sigma^0 p$ scattering from~\cite{lambdansigma0}.

\item {$\Sigma^++p\rightarrow \Sigma^+ +p$}

$\sigma =11.77/p_{LAB}+19.07$.

Fitted to data for 
$\Sigma^+ p\rightarrow \Sigma^+ p$ scattering from 
refs.~\cite{sigmap}.

\item {$\Sigma^-+p\rightarrow \Sigma^- +p$}

$\sigma =22.40/p_{LAB}-1.08$.

Fitted to data for 
$\Sigma^- p\rightarrow \Sigma^- p$ scattering from 
\cite{sigmap}.

The rest of the channels have not been fitted and we have used either isospin symmetry,
detailed balance or assumed a similar size and energy dependence to the available
channels.

\item {$\sigma_{\Lambda+n\rightarrow \Sigma^- p}\, =\, \sigma_{\Lambda+p\rightarrow 
\Sigma^+ n}\, = \, 2 \sigma_{\Lambda+n\rightarrow \Sigma^0+ n}
\, = \, 2 \sigma_{\Lambda+p\rightarrow \Sigma^0+ p}\,$,
$\sigma_{\Sigma^-+n\rightarrow \Sigma^- +n}\, =\, 
\sigma_{\Sigma^++p\rightarrow \Sigma^+ +p} $ and 
$\sigma_{\Sigma^++n\rightarrow \Sigma^+ +n}\, =\, 
\sigma_{\Sigma^-+p\rightarrow \Sigma^- +p} $} using isospin symmetry.

With these, we already have all channels with a $\Lambda$ in the initial state.
The missing channels with a $\Lambda$ in the final state are obtained by
detailed balance, so that 

$$p_{ab}^2 \sigma_{ab\rightarrow cd} =p_{cd}^2 \sigma_{cd\rightarrow ab}
$$

where $p_{ab}$ and $p_{cd}$ are the corresponding CM momenta. 
The rest of the $\Sigma+N$ processes have been taken with a cross section equal
to the $\Sigma^-+p\rightarrow \Sigma^- +p$. For the case 
$\Sigma^-+p\rightarrow \Sigma^0 +n $ there are a few data points
\cite{sigmampinel} compatible with this value.

\end{enumerate}

\end{document}